\documentclass[3p,times,twocolumn]{elsarticle}

\usepackage{ecrc}

\volume{00}

\firstpage{1}

\journalname{Nuclear Physics B Proceedings Supplement}

\runauth{}

\jid{nuphbp}

\jnltitlelogo{Nuclear Physics B Proceedings Supplement}

\usepackage{amssymb}

\usepackage[figuresright]{rotating}

\input babarsym

\begin{document}

\begin{frontmatter}



\dochead{}

\title{$e^+e^-$ results from BABAR and implications for the muon g-2}


\author{Michel Davier}

\address{LAL, CNRS/IN2P3 et Universit\'e Paris-Sud 11, 91898 Orsay, France}

\begin{abstract}
The \babar\ collaboration has nearly completed a program of precise 
measurements of the cross sections for the dominant channels of 
$\epem\to hadrons$ from threshold to an energy of $3-5\gev$ using 
the initial-state radiation (ISR) method, {\it i.e.} the measurement of the
cross sections $\epem\to \g~hadrons$ with the energetic $\g$ detected at 
large angle to the beams. These data are used as input to vacuum polarization 
dispersion integrals, in particular the hadronic contribution to
the muon $g-2$ anomaly. In addition to the recently measured $\pi^+\pi^-$
cross section, giving the dominant contibution, many multihadronic channels
have been investigated, with some recent examples presented here. We give 
preliminary results for the process $\epem\to\KpKm(\g)$ using $232\invfb$ of 
data collected with the \babar\ detector at $e^+e^-$ center-of-mass energies 
near $10.6\gev$. The lowest-order contribution to the hadronic vacuum 
polarization term in the muon magnetic anomaly is obtained for this channel: 
$a_\mu^{\rm KK, LO}=(22.95\pm0.14_{\rm(stat)}\pm0.22_{\rm (syst)})\times 10^{-10}$, 
which is about a factor of three more precise than the previous world average 
value. 
\end{abstract}

\begin{keyword}


\end{keyword}

\end{frontmatter}


\section{ Hadronic vacuum polarization and muon g-2}
\label{HVP-intro}
An important part of the Standard Model prediction for the muon magnetic 
anomaly $a_\mu=(g-2)/2$, where $g$ is the gyromagnetic ratio equal to 2 at 
lowest QED order, is given by hadronic vacuum polarization (HVP). In fact the 
dominant uncertainty in the prediction comes from the HVP contribution which is
computed through a dispersion relation using the experimental information on
the cross section for $e^+e^-\rightarrow {\rm hadrons}$, as the relevant energy
scale is too low for applying perturbative QCD. The HVP component is given by:
\begin{equation}
   a_\mu^{had}=\frac {1}{4\pi^3}\!\!
    \int_{4m_\pi^2}^\infty\!\!ds\,K(s)\,\sigma^{0}_{\rm hadrons}(s)~,
\end{equation}
where $K(s)$ is a QED kernel and $\sigma^{0}_{\rm hadrons}(s)$ the bare
cross section including final state radiation (FSR).

\section{The ISR method at \babar}
\label{ISR-method}
Unlike previous measurements that were done by energy scans, the 
analyses presented here use the ISR method~\cite{isr}. 
The $e^+e^- \rightarrow X(\gamma)$ cross section at the reduced 
energy $\sqrt{s'}$ is deduced from the measured spectrum
of $e^+e^- \to X(\gamma)\gamma_{\rm ISR}$ events produced at the
center-of-mass (c.m.) energy $\sqrt{s}$. The reduced energy is related to the
energy $E_\gamma^*$ of the ISR photon in the $e^+e^-$ c.m.\ frame by
$s'=s(1-2E_\gamma^*/\sqrt{s})$.
$\sqrt{s'}$ is equal to the mass of the final state $X$, including 
FSR photons. The ISR method follows from the relation
\begin{equation}
\label{Eq:def-lumi}
  \frac {dN_X(\gamma)\gamma_{\rm ISR}}{d\sqrt{s'}}\!=
   \!\frac {dL_{\rm ISR}^{\rm eff}}{d\sqrt{s'}}~
    \epsilon_{X\gamma}(\sqrt{s'})~\sigma_{X(\gamma)}^0(\sqrt{s'}),
\end{equation}
where $dL_{\rm ISR}^{\rm eff}/d\sqrt{s'}$ is the effective ISR luminosity,
$\epsilon_{X\gamma}$ is the full acceptance for the event sample, and
$\sigma_{X(\gamma)}^0$ is the `bare' cross section for the process $e^+e^-
\to X(\gamma)$ (including final-state radiative effects, but with leptonic 
and hadronic vacuum polarization contributions excluded). For precision
measurements with the ISR method, the effective ISR luminosity is not taken 
from the theoretical radiator function, which describes the probability 
to emit an ISR photon of energy $E_{\gamma}^*$ in a given angular acceptance, 
and the knowledge of the $e^+e^-$ luminosity. Instead, it 
is determined from the measurement of the
$e^+e^-\rightarrow \mu^+\mu^-(\gamma)\gamma_{\rm ISR}$ spectrum on the same 
data sample, through a relation similar to Eq.~(\ref{Eq:def-lumi}) where the 
$e^+e^- \to \mu^+\mu^-$ cross section is given by QED. In this way several
systematic uncertainties cancel. In particular, the measurement 
is mostly insensitive to higher order ISR corrections and other
uncertainties affecting the hadron and muon channels equally.

In the \babar\ analyses the ISR photon is detected at large angle with 
$E_\gamma^*>3$~GeV. This defines a topology where the ISR photon is 
back-to-back to the produced hadrons, thus providing high acceptance and
better particle identification (PID). Kinematic fits are
used to reject backgrounds and improve mass resolution. A continuous cross 
section measurement from threshold up to 3-5 GeV is achieved, the upper range 
value depending on the background for each exclusive process.

\section{Overview of \babar\ ISR results}
\label{babar-ISR}
So far \babar\ has published cross section results on 21 exclusive channels:
$\pi^+\pi^-$~\cite{prl-pipi,prd-pipi}, $2(\pi^+\pi^-)$~\cite{4pi2},
$\pi^+\pi^-\pi^0$, $K^+K^-\pi^+\pi^-$, $K^+K^-2\pi^0$, $2(K^+K^-)$,
$K_SK^\pm\pi^\mp$, $K^+K^-\pi^0$, $K^+K^-\eta$, $2(\pi^+\pi^-)\pi^0$, 
$2(\pi^+\pi^-)\eta$, $K^+K^-\pi^+\pi^-\pi^0$, $K^+K^-\pi^+\pi^-\eta$,
$3(\pi^+\pi^-)$, $2(\pi^+\pi^-\pi^0)$, $2(\pi^+\pi^-)K^+K^-$, 
$\phi\eta$, $\phi f^0(980)$, $p\overline{p}$, $\Lambda\overline{\Lambda}$, 
$\Lambda\overline{\Sigma^0}$ and c.c., $\Sigma^0\overline{\Sigma^0}$~\cite{babar-all}.

Some final states are still under study: $\pi^+\pi^-2\pi^0$, $K_SK_L$, 
$K_SK_L\pi^+\pi^-$, $K_SK^\pm\pi^\mp\pi^0$, $K_SK^\pm\pi^\mp\eta$.
Preliminary results on $K^+K^-$ are given in the next section.

\begin{figure}[t]
  \centering
  \vspace{-4.7cm}
  \includegraphics[width=9cm]{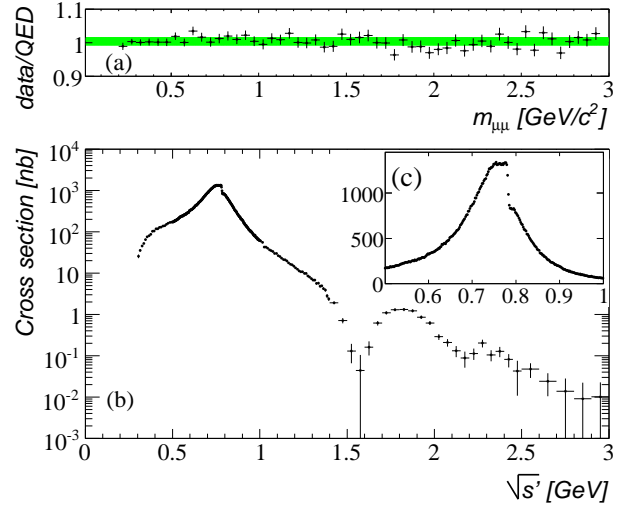}
  \vspace{-0.6cm}
  \caption{\small
(a) The ratio of the measured cross section for 
$e^+e^-\to\mu^+\mu^-\gamma(\gamma)$ to the NLO QED prediction. 
The band represents the best fit with the total uncertainty.
(b) The measured cross section for $e^+e^-\to\pi^+\pi^-(\gamma)$ 
from 0.3 to $3\gev$. 
(c) Enlarged view of the $\rho$ region in energy intervals of 2 MeV.
The plotted errors are from the sum of the diagonal elements of the 
statistical and systematic covariance matrices from the unfolding procedure.}
  \label{fig-pipi}
\end{figure}

Some recently published results can be highlighted in view of their importance
for $g-2$. The measurement of $\sigma (e^+e^-\rightarrow \pi^+\pi^-(\gamma)$
is the most precise and complete for this process~\cite{prl-pipi,prd-pipi}. 
The \babar\ ISR procedure is checked with the QED reaction 
$\sigma (e^+e^-\rightarrow \mu^+\mu^-(\gamma))$.
Both results are given in Fig.~\ref{fig-pipi}, showing agreement within 1.1\% 
(dominated by the 0.94\% uncertainty on the $e^+e^-$ luminosity) of the 
$\mu\mu(\gamma)$ data with the next-to-leading-order (NLO) QED calculation.
The $\pi\pi(\gamma)$ cross section measurement, being obtained from the ratio 
$\pi\pi/\mu\mu$, does not rely on the $e^+e^-$ luminosity, thus providing a
systematic uncertainty of only 0.5\% in the dominant $\rho$ region. 

\begin{figure}[t]
  \centering
  \vspace{-2.7cm}
  \includegraphics[width=7.6cm]{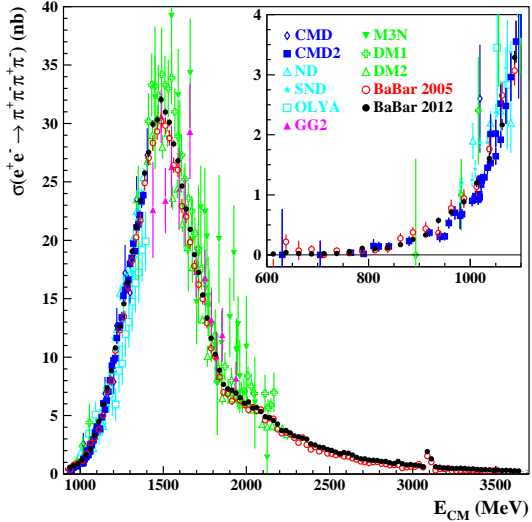}
  \vspace{-0.4cm}
  \caption{\small
The new \babar\ results (filled circles) on the 
$e^+e^-\rightarrow 2(\pi^+\pi^-)$ cross section in comparison to previous 
experiments. Only statistical uncertainties are shown.}
  \label{fig-4pi}
\end{figure}

Another important process is $e^+e^-\rightarrow 2(\pi^+\pi^-)$ on which an 
analysis based on the full \babar\ integrated luminosity of 454 fb$^{-1}$
has been recently 
published~\cite{4pi2}. The cross section results are given in 
Fig.~\ref{fig-4pi} together with those from previous experiments. Agreement
is reasonable, the \babar\ results being both more precise and spanning the
whole energy range of interest for HVP dispersion integrals.

\section{Preliminary results on $e^+e^-\rightarrow K^+K^-(\gamma)$}

\subsection{Event selection and analysis}
The analysis is based on $232\invfb$ of data collected by the \babar\ 
detector~\cite{detector} and follows closely the $\pi\pi$ 
analysis~\cite{prd-pipi}. Two-track ISR events are selected by requiring a 
photon with an energy $E_\gamma^*>3\gev$ in the $e^+e^-$ c.m.\ and polar angle 
with respect to the $e^-$ beam in the range [0.35--2.4]\rad, and exactly two 
tracks of opposite charge, each with momentum $p>1\gevc$ and identified as
kaons with the DIRC Cerenkov detector and $dE/dx$. A $K$-ID efficiency of 80\% 
is achieved, with fake rates at most 10\% at the highest momentum. Kaon ID, as
well as pion and muon mis-ID to kaons, are studied with high purity data 
samples.

As for the analysis of the $\mu\mu\gamma$ and $\pi\pi\gamma$ processes, the 
event definition is enlarged to include the radiation of one photon in 
addition to the already required ISR photon. Two kinematic fits to 
$e^+e^- \to K^+K^- (\gamma) \gamma_{\rm ISR}$ are performed: 
(1) if an additional 
photon is detected in the EMC, with energy $E_{\gamma}>20\mev$, it is 
used in a three-constraint (3C) fit, called `FSR' fit (however the 
extra photon can be either from FSR or additional ISR); 
(2) a fit with the additional photon assumed to be emitted along the 
$e^\pm$ beam directions, called 2C ISR fit. This procedure allows the 
reconstruction of events at NLO, therefore reducing the uncertainty resulting
from the neglect of events with higher-order radiation. Each event is 
characterized by the two $\chi^2_{\rm FSR}$ and $\chi^2_{\rm ISR}$ values 
(except for the 12.5\% with no extra measured photons), and the $KK$ mass 
is obtained from the ISR fit if $\chi^2_{\rm ISR}<\chi^2_{\rm FSR}$, and from 
the FSR fit in the reverse case. For the cross section measurement, the 
$KK(\gamma)\gamma_{\rm ISR}$ candidates are required to satisfy  
$\ln(\chi^2_{\rm ISR}+1)<3$.

The overall acceptance is determined with a full simulation, with corrections 
applied to account for observed data to MC differences.
Through specific studies, the ratios of efficiencies are obtained 
in data and simulation for trigger, tracking, PID and $\chi^2$ selection, 
and applied as mass-dependent corrections to the $KK$ mass spectrum measured
in data. Small corrections to the geometrical acceptance due to the assumed 
collinear additional ISR in the generator are obtained from the PHOKHARA 
code~\cite{phok} with fast simulation. 

Backgrounds stem primarily from other ISR events with extra $\pi^0$'s or 
misidentified pions, muons,or protons. Non-ISR $q\bar{q}$ represents the 
other important source of background, with an energetic photon from $\pi^0$ 
decay misidentified as the ISR photon. Estimation of backgrounds rely on
PID measurements, rescaled simulation results using comparison with data in
special samples, and the shape of the $\chi^2_{ISR}$ distribution. Background
levels are negligible on the $\phi$ resonance, but increase at larger masses
(about 2\% at 1.1 GeV, 20\% near 1.2 GeV from $\rho\rightarrow \pi\pi$, 10\%
at 2 GeV, 50\% at 5 GeV ). After specific studies of the mass
calibration and resolution, unfolding of the background-subtracted mass 
spectrum is performed. Finally the effective luminosity is obtained with the 
$\mu\mu(\gamma)$ sample.

\subsection{Results}
\begin{figure}[t]
  \centering
  \vspace{-5.9cm}
  \includegraphics[width=9.0cm]{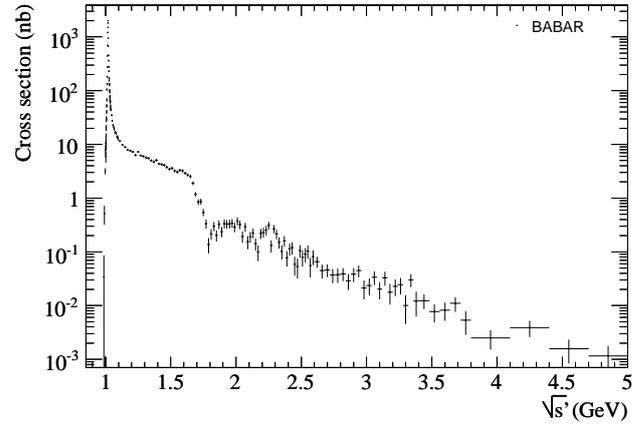}
  \vspace{-0.6cm}
  \caption{\small
The preliminary \babar\ $e^+e^- \to K^+K^-(\gamma)$ bare cross section 
including FSR. 
Systematic and statistical uncertainties are shown as 
diagonal  elements of the total covariance matrix. The contributions of
the $J/\psi$ and $\psi(2S)$ resonances have been removed on this plot.}
  \label{fig-KKxsec}
\end{figure}

The $\sigma_{K^+K^-(\gamma)}^0(\sqrt{s'})$ bare cross section  
including FSR is shown in Fig.~\ref{fig-KKxsec},
from threshold up to 5 GeV. It spans over a very large
dynamical range~(more than six orders of magnitude), and is dominated by the
$\phi$ resonance close to threshold. Other structures are clearly
visible at higher masses.

\begin{figure}[t]
  \centering
  \vspace{-5.9cm}
  \includegraphics[width=8.8cm]{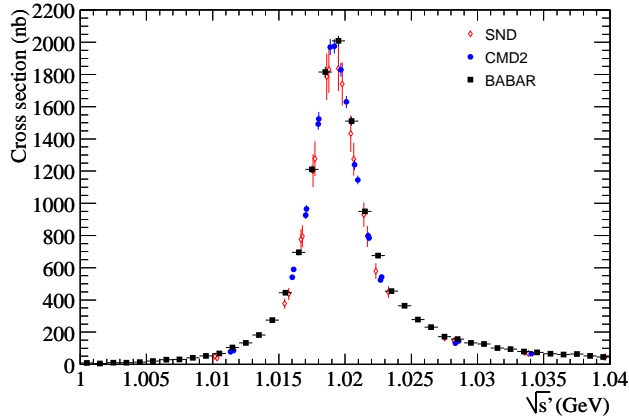}
  \vspace{-0.6cm}
  \caption{\small
The preliminary \babar\ $e^+e^- \to K^+K^-(\gamma)$ bare cross section 
including FSR in the $\phi$ region. Data points from previous
CMD-2 and SND experiments are shown for comparison.}
  \label{fig-phi}
\end{figure}

The cross section in the $\phi$ region is given in Fig.~\ref{fig-phi}. 
It is in fair agreement with previous CMD-2~\cite{cmd2} and SND~\cite{snd} s
results: the $\phi$ peak \babar\ cross section is
about 5\% (10\%) higher than CMD-2 (SND), where the quoted systematic 
uncertainties 
are 0.7\%, 2.2\%, 7.1\% for \babar\, CMD-2, SND, respectively. There is also
a small shift in mass (92 keV with CMD-2 and 65 keV with SND) consistent with
the quoted mass calibration uncertainties, 110 keV (a preliminary conservative
estimate) for \babar\ and 80 keV for CMD-2). 

A fit of the charged kaon form factor measured by \babar\ is performed taking
into account the $\rho$ and $\omega$ tails, as well as contributions from 
higher mass vector bosons. The following $\phi$ parameters are obtained:
$m_\phi=(1019.51 \pm 0.02_{\rm exp} \pm 0.11_{\rm cal})~{\rm MeV}$, 
$\Gamma_\phi=(4.29 \pm 0.04_{\rm exp} \pm 0.07_{\rm resol})~{\rm MeV}$, 
$\Gamma_{\rm ee}^{\phi}  B^\phi_{K^+K^-}=
(0.6344\pm 0.0059_{\rm exp} \pm 0.0028_{\rm fit} \pm 0.0015_{\rm cal})~{\rm keV}$,
where the last two uncertainties are from the form factor fit and the mass
calibration. The precision of the last result is improved by a factor of two
compared to the best value from CMD-2~\cite{cmd2}.
 
\section{The impact of \babar\ data on the $g-2$ prediction}

\begin{figure}[t]
  \centering
  \vspace{-1.8cm}
  \includegraphics[width=6.9cm]{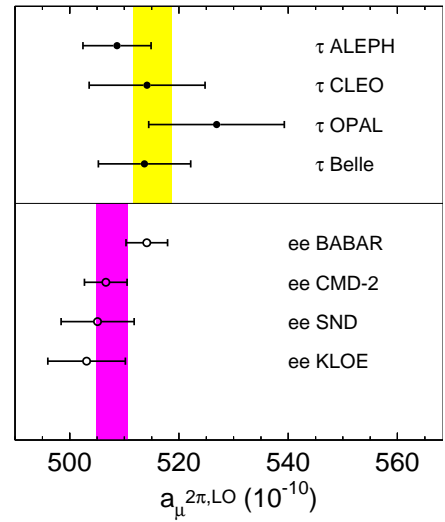}
  \vspace{-0.4cm}
  \caption{\small
The $\pi\pi$ HVP contribution to $a_\mu$ obtained from $\tau$ decays with
isospin-breaking corrections (top) and
$e^+e^-$ data (bottom)~\cite{newtau,newee}.}
  \label{fig-amu-2pi}
\end{figure}

The dominant $\pi\pi$ HVP contribution to $a_\mu$ has received a lot of 
attention over the last 10 years, with discrepancies between experiments 
being partially resolved with time. Fig.~\ref{fig-amu-2pi} shows the results
from all experiments as well as the determinations using $\tau$ data corrected
for isospin-breaking~\cite{newtau}. To keep results as independent as possible
the comparison uses only data from the considered experiment (complemented by 
world-average data in energy ranges not covered). It is seen that the \babar\ 
result is the most precise (with CMD-2) and helps reduce the tension between 
$ee$ and $\tau$ data.

The precision of the new data presented here on the $K^+K^-$ and 
$2(\pi^+\pi^-)$ channels also allow further progress. The contributions up to 
1.8 GeV are 
$a_\mu^{K^+K^-,\rm LO} = \left(22.95 \pm 0.14_{\rm stat} \pm 0.22_{\rm syst}\right)\times10^{-10}$ and 
$a_\mu^{2(\pi^+\pi^-),\rm LO} = \left(13.64 \pm 0.03_{\rm stat} \pm 0.36_{\rm syst}\right)\times10^{-10}$, respectively
to be compared with previous determinations~\cite{dhmz-last}, 
$\left(21.63 \pm 0.27_{\rm stat} \pm 0.68_{\rm syst}\right)\times10^{-10}$ and 
$\left(13.35 \pm 0.10_{\rm stat} \pm 0.52_{\rm syst}\right)\times10^{-10}$. 
For other multihadronic
channels the \babar\ ISR results are by far the most accurate and complete. 
In addition the dynamics of each final state has been studied and found 
to be dominated by resonances which have been identified. This is 
important because it is possible in this way to derive some cross section 
estimates~\cite{dhmz-last} for final states which are difficult to measure 
($> 2\pi^0$) using the known branching fractions of these resonances. This 
detailed information has been used in the most recent estimate of $a_\mu$,
showing a $3.6\sigma$ discrepancy~\cite{dhmz-last} with the direct 
measurement~\cite{bnl}. Another estimate ~\cite{hlmnt} uses also \babar\ 
cross section data.

I would like to thank the Nagoya group for organizing a perfect $\tau$ 
workshop.





\begin{thebibliography}{00}

\bibitem{isr}     V.N. Baier and V.S. Fadin, 
                           Phys.Lett. B27, 233~(1968);    
                  A.B. Arbuzov {\it et al.}, 
                           J. High Energy Phys. {\bf 9812}, 009 (1998);
                  S. Binner, J.H. K\"uhn, and K. Melnikov, 
                           Phys.Lett. {\bf B459}, 279 (1999);
                  M. Benayoun {\it et al.}, 
                           Mod.Phys.Lett. {\bf A14}, 2605 (1999).
\bibitem{prl-pipi} B. Aubert {\it et al.}, \babar\ Collaboration, Phys.Rev.Lett. {\bf 103}, 231801 (2009).
\bibitem{prd-pipi} J.P. Lees {\it et al.}, \babar\ Collaboration, Phys.Rev. {\bf D86}, 032013 (2012).
\bibitem{4pi2} J.P. Lees {\it et al.}, \babar\ Collaboration, Phys.Rev. {\bf D85}, 112009 (2012).
\bibitem{babar-all}   J.P. Lees {\it et al.}, \babar\ Collaboration,
                     Phys.Rev. {\bf D86}, 012008 (2012);
                      B. Aubert {\it et al.}, \babar\ Collaboration, 
                     Phys.Rev. {\bf D77}, 092002 (2008);
                     Phys.Rev. {\bf D76}, 012008 (2007);
                     Phys.Rev. {\bf D76}, 092006 (2007);
                     Phys.Rev. {\bf D76}, 092005 (2007);
                     Phys.Rev. RC {\bf D74}, 091103 (2006);
                     Phys.Rev. RC {\bf D74}, 111103 (2006);
                     Phys.Rev. {\bf D73}, 052003 (2006);
                     Phys.Rev. {\bf D73}, 012005 (2006);
                     Phys.Rev. {\bf D71}, 052001 (2005);
                     Phys.Rev. {\bf D70}, 072004 (2004).
\bibitem{detector}    B. Aubert {\it et al.},
                           Nucl.Instr.Meth. {\bf A479}, 1 (2002).
\bibitem{phok}    H. Czy\.z {\it et al.},
                           Eur.Phys.J. {\bf C35}, 527 (2004); 
                           Eur.Phys.J. {\bf C39}, 411 (2005).
\bibitem{cmd2}     R.R. Akhmetshin {\it et al.}, 
                           Phys.Lett. {\bf B669}, 217-222 (2008). 
\bibitem{snd}      M.N. Achasov {\it et al.}, 
               Phys.Rev. {\bf D63}, 072002 (2001);
               Phys.Rev. {\bf D76}, 072012 (2007).
\bibitem{newtau}      M. Davier {\it et al.},  
                           Eur. Phys. Jour. {\bf C66}, 127 (2010).
\bibitem{newee}   M. Davier {\it et al.},
		          Eur.Phys.J. {\bf C66}, 1 (2010).
\bibitem{dhmz-last} M. Davier, A. Hoecker, B. Malaescu, and Z. Zhang,
   Eur.Phys.J. {\bf C71}, 1515 (2011), Erratum-ibid. {\bf C72} 1874 (2012).
\bibitem{bnl}         G.W. Bennett {\it et al.},
                           Phys. Rev. {\bf D73}, 072003 (2006).
\bibitem{hlmnt}   K.~Hagiwara {\it et al.}, 
                           J. Phys. {\bf G 38}, 085003 (2011).



\end{thebibliography}



\end{document}